\documentclass[final]{siamltex}

\usepackage{graphicx}

\usepackage{latexsym}
\usepackage{amsmath}
\usepackage{amsfonts}
\usepackage{amssymb}
\usepackage{amsbsy}

\usepackage{natbib}
\bibpunct{(}{)}{,}{a}{,}{,}

\newcommand\EI{{\em Error and Inference}}

\newenvironment{myquote}
   {\smallskip\begin{quote}\small\begin{em}}
   {\normalsize\end{em}\end{quote}\smallskip}
\newcommand\beq{\begin{myquote}}
\newcommand\enq{\end{myquote}}

\title{Error and Inference:\\ 
an outsider stand\\ 
on a frequentist philosophy}

\author{Christian P. Robert\\ 
Universit{\'e} Paris-Dauphine \\
CEREMADE, IUF, and CREST, \\
75775 Paris cedex 16, France \\
{\sf xian@ceremade.dauphine.fr}
}
\begin{document}

\maketitle

\begin{abstract}
This note is an extended review of the book \EI, edited by Deborah Mayo and Aris Spanos, about
their frequentist and philosophical perspective on testing of hypothesis and on the criticisms of alternatives 
like the Bayesian approach.
\end{abstract}

\pagestyle{myheadings}
\thispagestyle{plain}

\section{Introduction}
\label{sec:intro}

    \begin{myquote}``The philosophy of science offer valuable tools for understanding and advancing solutions to the problems
of evidence and inference in practice"{\em ---D. Mayo and A. Spanos, p.xiv, \EI, 2010}\end{myquote}

This philosophy book, \EI, whose subtitle is {\em ``Recent exchanges on experimental reasoning, reliability,
and the objectivity and rationality of Science"}, includes contributions by P.~Achinstein, A.~Chalmers, D.~Cox,
C.~Glymour, L.~Laudan, A.~Musgrave, and J.~Worrall, and by both editors, D.~Mayo and A.~Spanos. It offers
reflections on the nature of testing and the defence of the ``Error-Statistical philosophy", proposed by
\cite{mayo:1996}. \EI~is the edited continuation of a conference held at Virginia Tech in 2006, {\em ERROR 06}.
Given my layman interest in the philosophy of Science, I find that the book reads rather well, despite my
missing references in the field. (Paradoxically, the sections I found the hardest to follow were those most
centred on Statistics.) The volume is very carefully edited and thus much more than a collection of papers,
thanks to the dedications of the editors. In particular, Deborah Mayo added her comments at the end of the
chapters of all contributors (but her own's and Aris Spanos'). Her strongly frequentist perspective on the
issues of testing and model choice are thus reflected in the tone of the volume, even though some contributors
bring some level of (un-Bayesian) contradiction to the debate. My own Bayesian arguments following from reading
\EI~are provided in the following sections, ordered by book chapters.  A broader and deeper
philosophico-statistical perspective on the nature of Bayesian inference to which I mostly subscribe are
given by \cite{berger:2003} and \cite{gelman:shalizi:2012}.

\section{Severe testing}\label{sec:sever}

    \beq``However, scientists wish to resist relativistic, fuzzy, or post-modern turns (...) Notably, the Popperian
requirement that our theories are testable and falsifiable is widely regarded to contain important insights
about responsible science  and objectivity."{\em --- D. Mayo and A. Spanos, p.2, \EI, 2010}\enq

As illustrated by the above quote (which first part I obviously endorse), the overall perspective in the book
is Popperian, despite Popper's criticism of statistical inference as a whole and of Bayesian statistics as a
particular (see the quote at the top of Section \ref{sec:cox+mayo}). Another fundamental concept throughout the
book is the ``Error-Statistical philosophy" whose Deborah Mayo is the proponent \citep{mayo:1996}. One of the tenets of this
philosophy is a reliance on statistical significance tests in the fusion Fisher-Neyman-Pearson (or
self-labelled {\em frequentist}) tradition, along with a severity principle (``We want hypotheses that will allow
for stringent testing so that if they pass we have evidence of a genuine experimental effect", p.19) stated as
(p.22)
\begin{quote}{\em 
    A hypothesis $H$ passes a severe test $T$ with data x if
    \begin{enumerate}
       \item $x$ agrees with $H$, and
       \item with very high probability, test $T$ would have produced a result that accords less well with $H$ than
	does $x$, if $H$ were false or incorrect.
    \end{enumerate}
}\end{quote}
The $p$-value is advanced as a direct accomplishment of this goal, but I fail to see why it does or, if it
does, then why a Bayes factor \citep{jeffreys:1939} would not. Indeed, the criterion depends on the definition
of the probability model when $H$ is false or incorrect.  This somehow reflects negatively Mayo's criticism of
the Bayesian approach, as explained below. In particular, the ``high probability" assumes a lack of contiguity
between the tested hypothesis $H$ and its negative $\neg H$, that is unrealistic even within the perspective of
the book, as is the claim below. (See also Spanos' required ``highly improbable" good fit to the model defined
by $H$ ``were $H$ to be in error", p.209.) 

    \beq``Formal error-statistical tests provide tools to ensure that errors will be correctly detected with high
probabilities." {\em --- D. Mayo, p.33, \EI, 2010}\enq

In Chapter 1 ({\em Learning from error, severe testing, and the growth of theoretical knowledge}), Deborah Mayo
address a frontal attack against the Bayesian approach. The main criticism therein is about the Bayesian approach
to testing (restrictedly defined through the posterior probability of the hypothesis, rather than through the
predictive) is about the ``catchall hypothesis", a somehow desultory term replacing the more standard
``alternative" hypothesis.  According to Mayo, this alternative hypothesis should ``include all possible rivals, including 
those not even though of" (p.37). This sounds to me like a weak argument, given that 
\begin{enumerate}
\item the argument should
also apply in the frequentist case, in order to define the relevant probability distribution ``when $H$ is false or
incorrect" (see, e.g., the above quote and also the one about the ``probability of so good an agreement (between $H$ and 
$x$) calculated under the assumption that $H$ is false", p.40); 
\item it is reasonable to argue that a well-defined alternative should always be available given testing an hypothesis is 
very rarely the ultimate goal of a study: if $H$ is rejected, there should be a available alternative model for
conducting the analysis of the available data, to be picked or constructed precisely among those ``thought of";
\item the argument about the infinite set of possible rivals is self-defeating in that it leads to the fallacy
of the perfect model: given a dataset, there always is a model that fits perfectly this dataset;
\item rejecting or accepting an hypothesis $H$ in terms of the sole null hypothesis $H$ does not make sense from 
operational as well as from game-theoretic \citep{degroot:1970} perspectives. 
\end{enumerate}
A further argument found in this chapter that the posterior probability of $H$ is a direct function of the
prior probability of $H$ does not stand when used against the Bayes factor. (The same lack of justification
applies to the criticism that the Bayesian approach does not accommodate newcomers, i.e., new alternatives,
since marginal likelihoods are computed separately for each potential model, \citealp{kass:raftery:1995}.)
Stating that ``one cannot vouch for the reliability of [this Bayesian] procedure---that it would rarely affirm
theory $T$ were $T$ false" (p.37) completely ignores the wealth of results about the consistency of the Bayes
factor \citep{schervish:1995,berger:ghosh:mukhopadhyay:2003,moreno:giron:casella:2010}, since the ``asymptotic
long run" (p.20) matters in the Error-Statistical philosophy. The final argument that Bayesians rank ``theories
that fit the data equally well (i.e., have identical likelihoods)" (p.38) does not account for---or dismisses,
see page 50 referring to Jeffreys and Berger instead of \cite{berger:jefferys:1992}---the fact that Bayes
factors are automated Occam's razors in that the averaging of the likelihoods over spaces of different
dimensions are natural advocates of simpler models \citep{mackay:2002}. Even though I discuss this issue in the
following section, Mayo also seems to imply there that Bayesians are using the data twice (this is at least how
I interpret the insistence on ``same" p. 50), which is a sin [genuine] Bayesian analysis can hardly be found
guilty of! Overall, the undercurrents of those criticisms is similar to the ones found in the recent attacks of
\cite{templeton:2008,templeton:2010}, later rebuked in, e.g., \cite{clade:2010}, about apparently the
approximative Bayesian calculation methods but in truth directed to the overall Bayesian handling of testing of
hypotheses.

\section{Theory confirmation and model evidence (Chapter 4)}
     \beq``Taking the simple case where the background principles specify a finite list of alternatives
$T_1,\ldots,T_n$, each piece of data falsifies some $T_i$ until we are left with just one theory $T_j$---which
(...) is thus `deduced from the phenomenon'.'" {\em --- J. Worrall, p.134, \EI, 2010}\enq

The fourth chapter of \EI, written by John Worrall, covers the highly interesting issue of ``using the data
twice". The point has been debated many times in the Bayesian literature and this is one of the main criticisms
raised against Aitkin's (\citeyear{aitkin:2010}) integrated likelihood. Worrall's perspective is both related
and unrelated to this purely statistical issue, when he considers that ``you can't use the same fact twice,
once in the construction of a theory and then again in its support" (p.129). (He even signed a ``UN Charter",
where UN stands for ``use novelty"!) After reading both Worrall's and Mayo's viewpoints, the later being that
all that matters is severe testing as it encompasses the UN perspective, I afraid I am none the wiser about the
relevance of their arguments from a statistical perspective, but this led me to reflect on the statistical
issue.\footnote{However, the UN principle seems much too mechanical when considering the quote from John
Worrall: ticking models away by using one observation at a time is certainly not a sound statistical procedure,
even though sufficient and testing statistics may vary from one model to the next
\citep{robert:cornuet:marin:pillai:2011}.}

From first principles, a Bayesian approach should only use the data once, namely when constructing the
posterior distribution on every unknown component of the model(s).  Given this all-encompassing posterior, all
inferential aspects are the consequences of a sequence of decision-theoretic steps in order to select optimal
procedures. This is the ideal setting while, in practice, relying on a sequence of posterior distributions is
often necessary, each posterior being a consequence of earlier decisions, which makes it the result of a
multiple use of the data... For instance, the process of Bayesian variable selection is on principle clean from
the sin of ``using the data twice": one simply computes the posterior probability of each of the variable
subsets and this is over. However, in a case involving many (many) variables, there are two difficulties: one
is about building the prior distributions for all possible models, a task that needs to be automatised to some
extent; another is about exploring the set of potential models, e.g., resorting to projection priors as in the
intrinsic solution of \cite{perez:berger:2002}, while unavoidable and a ``least worst" solution, means
switching priors/posteriors based on earlier acceptances/rejections, i.e.~on the data. Second, the path of
models truly explored [which will be a minuscule subset of the set of all models] will depend on the models
rejected so far, either when relying on a stepwise exploration or when using a random walk MCMC algorithm.
Although this is not crystal clear (there actually is plenty of room for arguing the opposite!), it could be
argued that the data is thus used several times in this process...

    \beq``Although the [data] set as a whole both fixes parameter values and (unconditionally) supports, no
particular element of the data set does both."{\em --- J. Worrall, p.140, \EI, 2010}\enq

One paragraph in Worrall's chapter intersects with the previous discussion, while getting away from the ``using
the data twice" discussion. It compares two theories with a different number of ``free" parameters, hence
(seemingly) getting different amounts of support from a given dataset (``$n$ lots of confirmation [versus]
$n-r$ lots", p.140). The distinction sounds too arithmetic in that the algebraic dimension of a parameter may
be only indirectly related to its information dimension, as illustrated by DIC \citep{spiegbestcarl}, although
this criterion does indeed use the data twice. Furthermore, a notion like the fractional Bayes factor
\citep{ohagan:1995} shows that the whole dataset may contribute both to the model selection and to the
parameter estimation without the above dichotomy to occur.

\section{Theory testing in economics and the error-statistical perspective (Chapter 6)}
    \beq``Statistical knowledge is independent of high-level theories." {\em --- A. Spanos, p.242, \EI, 2010}\enq

The sixth chapter of \EI~is written by Aris Spanos and deals with the issues of testing in econometrics (rather
than economics). It provides on the one hand a fairly interesting entry in the history of economics and the
resistance to data-backed theories, primarily because the buffers between data and theory are multifold (``huge
gap between economic theories and the available observational data``, p.203). On the other hand, what I fail to
understand in the chapter (and in other parts of \EI) is the (local) meaning of theory, as it seems very
distinct from what I would call a (statistical) model. The sentence ``statistical knowledge, stemming from a
statistically adequate model allows data to `have a voice of its own's (...) separate from the theory under
scrutiny and its succeeds in securing the frequentist goal of objectivity in theory testing" (p.206) is
puzzling in this respect. (Actually, I would have liked to see a clear meaning put to this ``voice of its own",
as it otherwise sounds mostly as a catchy sentence...) Similarly, Spanos distinguishes between three types of
models: primary/theoretical, experimental/structural: ``the structural model contains a theory's substantive
subject matter information in light of the available data" (p.213), data/statistical: ``the statistical model
is built exclusively using the information contained in the data" (p.213). I have trouble to understand how
testing can distinguish between those types of models: as a na{\"\i}ve reader, I would have thought that only
the statistical model could be tested by a statistical procedure, even though I would not call the above a
proper definition of a statistical model (esp. since Spanos writes a few lines below that the statistical model
``would embed (nest) the structural model in its context" (p.213)). The normal example followed on pages
213-217 does not help [me] to put meaning on this distinction: it simply illustrates the impact of failing some
of the defining assumptions (normality, time homogeneity [in mean and variance], independence). (As an aside,
the discussion about the poor estimation of the correlation p.214-215 does not help, because it involves a
second variable $Y$ that is not defined for this example.) It would be nice of course if the ``noise" in a
statistical/econometric model could be studied in complete separation from the structure of this model, however
they seem to be irremediably intermingled to prevent this partition of roles. I thus do not see how the
``statistically adequate model is independent from the substantive information" (p.217), i.e. by which rigorous
process one can isolate the ``chance" parts of the data to build and validate a statistical model per se. The
simultaneous equation model (SEM, pp.230-231) is more illuminating of the distinction set by Spanos between
structural and statistical models/parameters, even though the difference in this case boils down to a question
of identifiability.

    \beq``What is needed is a methodology of error inquiry that encourages detection and identification of the
different ways an inductive inference could be in error by applying effective procedures that would detect such
errors when present with very high probability." {\em --- A. Spanos, p.241, \EI, 2010}\enq

The chapter, in line with the book, is strongly entrenched within the ``F-N-P" frequentist paradigm" (p.210).
Obviously, there are major differences between the Fisherian and the Neyman-Pearson approaches to testing that
are not addressed in the chapter, the main opposition being the role (or non-role) of the $p$-value. The
recurrent (and relevant) worry of Spanos about model misspecification is not directly addressed by either of
those. The extremely strong criticisms of ``cookbook" econometrics textbooks (p.233) could thus be equally
addressed to most statistics books and papers: I do not see how the ``error statistical perspective" could be
able to spot all departures from model assumptions. Section 6.3 comparing covariate dependent models \`a la
Cowles Commission with standard autoregressive models is thus puzzling because (a) they do not seem
particularly comparable to me, for the very reason evacuated by Spanos that ``ARIMA models ignore all
substantive information", and (b) time series models may be just as well misspecified. To think that adding a
linear time-dependence to a regression model is sufficient to solve the issues, as argued by Spanos  (``...what
distinguishes [this] approach from other more data-oriented traditions is its persistent emphasis on justifying
the methodological foundations of its procedures using the scientific credentials of frequentist inference``,
p.238), is a rather radical shortcut for a justification of the approach.

\section{New perspectives on (some old) problems of frequentist statistics (Chapter 7)}
    \beq``‘The defining feature of an inductive inference is that the premises (evidence statements) can be true
while the conclusion inferred may be false without a logical contradiction: the conclusion is ``evidence
transcending"." {\em --- D. Mayo and D. Cox, p.249, \EI, 2010}\enq

The seventh chapter of \EI~is divided in four parts, written by David Cox, Deborah Mayo, and Aris Spanos, in
different orders and groups of authors. This is certainly the most statistical of all chapters, not a surprise
when considering that David Cox was involved.  Overall, this chapter is crucial by its contribution to
the debate on the nature of statistical testing.

\subsection{Frequentist statistics as a theory of inductive inference (Part I)}\label{sec:cox+mayo}
    \beq``The advantage in the modern statistical framework is that the probabilities arise from defining a
probability model to represent the phenomenon of interest. Had Popper made use of the statistical testing ideas
being developed at around the same time, he might have been able to substantiate his account of
falsification." {\em --- D. Mayo and D. Cox, p.251, \EI, 2010}\enq

The first part of the chapter is Mayo's and Cox theory of inductive inference.  It was first published in the
2006 Erich Lehmann symposium \citep{mayo:cox:2006}. Contrary to Part II, there is absolutely no attempt there
to link nor clash with the Bayesian approach: this part is focussing on frequentist statistical theory as the
sole basis for inductive inference. The debate therein about deducing that $H$ is correct from a dataset
successfully facing a statistical test is classical (in both senses) but I stand unmoved by the arguments. The
null hypothesis $H$ remains the calibrating distribution throughout the chapter, with very little (or at least
not enough) consideration of what happens when the null hypothesis does not hold.  Section 3.6 [of this part]
about confidence intervals being another facet of testing hypotheses is representative of this perspective. The
$p$-value is defended as the central tool for conducting hypothesis assessment. (In this version of the paper,
some $p$'s are written in roman characters and others in italics, which is a wee confusing until one realises
that this is a mere typo!)
The fundamental imbalance problem, namely that, in contiguous hypotheses, a test cannot be expected both to
most often reject the null $H$ when it is [very moderately] false {\em and} to most often accept the null $H$
when it is correct is not discussed there. As stated in the introduction, the argument about substantive nulls
found in Section 3.5 applies to a stylised case of well-separated scientific theories, however the real world
of models is more similar to a grayish (and more Popperian?) continuum of possibles. In connection with this, I
would have expected that the book addressed on philosophical grounds George Box's aphorism that ``all models
are wrong". Indeed, one (both philosophical and methodological) difficulty with both the $p$-values and the
frequentist evidence principle (FEV) is that they rely on the strongly restrictive belief that one given model
can be exact or true (while criticising the subjectivity of the prior modelling in the Bayesian approach). Even
in the typology of types of null hypotheses drawn by the authors in Section 3, the ``possibility of model
misspecification" is addressed in terms of the low power of an omnibus test, while agreeing that ``an
incomplete probability specification" is unavoidable. An argument found at several place in \EI~is that the
alternative to $H$ cannot be completely specified, an argument going against the second part of Box's aphorism
that ``some models are more useful than others". 

    \beq``Sometimes we can find evidence for $H_0$, understood as an assertion that a particular discrepancy, flaw, or
error is absent, and we can do this by means of tests that, with high probability, would have reported a
discrepancy had one been present."{\em --- D. Mayo and D. Cox, p.255, \EI, 2010}\enq

The above quote relates to the {\em Failure and Confirmation} section where the authors try to push the
argument in favour of frequentist tests one step further, namely that that ``moderate $p$-values" may sometimes
be used as confirmation of the null. (I may have misunderstood, the end of the section appearing as a defence
of a purely frequentist---as in repeated experiments---interpretation. This reproduces an earlier argument
about the nature of probability found in Section 1.2, as characterising the ``stability of relative frequencies
of results of repeated trials".) 

As an aside, this chapter made me ponder afresh about the nature of probability, a debate that put me off so
much in \cite{keynes:1921} and even in \cite{jeffreys:1939}
\citep[see][]{robert:chopin:rousseau:2009,robert:2010}.  From a mathematical formal perspective, there is only
one ``kind" of probability, the one defined via a reference measure and a probability, whether it applies to
observations or to parameters. From a philosophical perspective, there is a natural issue about the ``truth" or
``realism" of the probability quantities and of the probabilistic statements. \EI~and in particular this
chapter consider that a truthful probability statement is the one agreeing with ``a hypothetical long-run of
repeated sampling, an error probability", while the statistical inference school of Keynes (1921), Jeffreys
(1939), and \cite{carnap:1952} ``involves quantifying a degree of support or confirmation in claims or
hypotheses", which makes this (Bayesian) approach sound less realistic. Obviously, I have no ambition to solve
this long-standing debate, however I see no reason in the first approach to be more realistic by being grounded
on stable relative frequencies \`a la \cite{vonmises:1957}.  If nothing else, the notion that a test should be
evaluated on its long run performances is very idealised as the concept relies on an ever-repeating and
infinite sequence of identical trials. Relying on probability measures as self-coherent mathematical measures
of uncertainty carries (for me) as much (or as little) reality as the above infinite experiment. Now, the paper
is not completely entrenched in this interpretation, e.g.~when it concludes that ``what makes the kind of
hypothetical reasoning relevant to the case at hand is not the long-run low error rates associated with using
the tool (or test) in this manner; it is rather what those error rates reveal about the data generating source
or phenomenon" (p.273).

    \beq``If the data are so extensive that accordance with the null hypothesis implies the absence of an effect of
practical importance, and a reasonably high $p$-value is achieved, then it may be taken as evidence of the
absence of an effect of practical importance."{\em --- D. Mayo and D. Cox, p.263, \EI, 2010}\enq

The paper discusses on several occurrences conclusions to be drawn from a $p$-value near one, as in the above
quote. This is an interpretation that does not sit well with my (and others',
e.g.~\citealp{hwang:casella:robert:wells:farrel:1992}) understanding of $p$-values being distributed as
uniforms under the null: very high  $p$-values should be as suspicious as very low $p$-values. (The criticism
is not new, of course, see \citealp{jeffreys:1939}.) Unless one does not strictly adhere to the null model,
which brings back the above issue of the approximativeness of any model. I also found fascinating to read the
criticism that ``power appertains to a prespecified rejection region, not to the specific data under analysis"
as I thought this equally applied to the $p$-values, turning ``the specific data under analysis" into a
departure event of a prespecified kind. (And leading to the famous aphorism by Jeffreys, 1939, about an
``hypothesis that may be true may be rejected because it has not predicted observable results that have not
occurred".)

\subsection{Objectivity and conditionality in frequentist inference (Part II)}
    \beq``Frequentist methods achieve an objective connection to hypotheses about the data-generating process by
being constrained and calibrated by the method's error probabilities in relation to these models ."{\em ---D. Cox and
D. Mayo, p.277, \EI, 2010}\enq

The second part of the seventh chapter of \EI, is due to David Cox and Deborah Mayo. The purpose
is clear and the sub-chapter quite readable from a statistician's perspective. I however find it difficult to
quantify objectivity by first conditioning on ``a statistical model postulated to have generated data", as again
this assumes the existence of a ``true" probability model where ``probabilities (...) are equal or close to  the
actual relative frequencies". As earlier stressed by Andrew Gelman on his blog,

    \beq``I don't think it's helpful to speak of ``objective priors." As a scientist, I try to be objective as much
as possible, but I think the objectivity comes in the principle, not the prior itself. A prior distribution---any
statistical model---reflects information, and the appropriate objective procedure will depend on what information
you have."{\em Andrew Gelman}\enq

This part opposes the likelihood, Bayesian, and frequentist methods, as in what \cite{gigerenzer:2002}
classifies as the ``superego, the ego, and the id". Cox and Mayo stress from the start that the frequentist
approach is (more) objective because it is based on the sampling distribution of the test.  My primary problem
with this thesis is that the ``hypothetical long run" (p.282) does not hold in realistic settings. Even in the
event of a reproduction of similar or identical tests, a sequential procedure exploiting everything that has
been observed so far is more efficient than the mere replication of the same procedure solely based on the
current observation.

    \beq``Virtually all (...) models are to some extent provisional, which is precisely what is expected in the
building up of knowledge."{\em ---D. Cox and D. Mayo, p.283, \EI, 2010}\enq

The above quote is something I completely agree with, being another phrasing of George Box's ``all models are
wrong", but this transience of working models is a good reason in my opinion to account for the possibility of
alternative working models from the start of the statistical analysis. Hence for an inclusion of those models
in the statistical analysis equally from the start. Which leads almost inevitably to a Bayesian formulation of
the testing problem.

    \beq``Perhaps the confusion [over the role of sufficient statistics] stems in part because the various
inference schools accept the broad, but not the detailed, implications of sufficiency."{\em ---D. Cox and D. Mayo,
p.286, \EI, 2010}\enq

The discussion over the sufficiency principle is quite interesting, as always. The authors propose to solve the
confusing opposition between the sufficiency principle and the frequentist approach by assuming that inference
``is relative to the particular experiment, the type of inference, and the overall statistical approach"
(p.287).  This constraint creates a barrier between sampling distributions that avoids the ``binomial versus
negative binomial" paradox often advanced in the Bayesian literature \citep{berger:wolpert:1988}. But the
authors' solution is somehow tautological: by conditioning on the sampling distribution, it avoids the
difficulties linked with several sampling distributions all producing the same likelihood.\footnote{From a
practical perspective, I have however become less enamoured of the sufficiency principle as the existence of
[non-trivial] sufficient statistics is quite the rare event. Especially across models, as discussed in
\cite{robert:cornuet:marin:pillai:2011}.} The section (pp.288-289) is also revealing about the above
``objectivity" of the frequentist approach in that the derivation of a test taking large value away from the
null with a well-known distribution under the null is not an automated process, esp. when nuisance parameters
cannot be escaped from (pp.291-294). Achieving separation from nuisance parameters, i.e. finding statistics
that can be conditioned upon to eliminate those nuisance parameters, does not seem feasible outside
well-formalised models related with exponential families.  Even in such formalised models, a (clear?) element
of arbitrariness is involved in the construction of the separations, which implies that the objectivity is
under clear threat. The chapter recognises this limitation in Section 9.2 (pp.293-294), however it argues that
separation is much more common in the asymptotic sense and opposes the approach to the Bayesian averaging over
the nuisance parameters, which ``may be vitiated by faulty priors" (p.294). I am not convinced by the argument,
given that the (approximate) condition approach amount to replace the unknown nuisance parameter by an
estimator, without accounting for the variability of this estimator. Averaging brings the right (in a
consistency sense) penalty.

A compelling section is the one about the weak conditionality principle (pp.294-298), as it objects to the
usual statement that a frequency approach breaks this principle. In a mixture experiment about the same
parameter $\theta$, inferences made conditional on the experiment  ``are appropriately drawn in terms of the
sampling behaviour in the experiment known to have been performed" (p. 296). This seems hardly objectionable, as
stated.  And I must confess the sin of stating the opposite as {\em The Bayesian Choice} has this remark
(\cite{robert:2007}, Example 1.3.7, p.18) that the classical confidence interval averages over the experiments.
The term experiment validates the above conditioning in that several experiments could be used to measure
$\theta$, each with a different $p$-value. I will not argue with this: I could, however, about ``conditioning
is warranted to achieve objective frequentist goals" (p. 298) in that the choice of the conditioning, among
other things, weakens the objectivity of the analysis. In a sense the above pirouette out of the conditioning
principle paradox suffers from the same weakness, namely that when two distributions characterise the same data
(the mixture and the conditional distributions), there is a choice to be made between ``good" and ``bad".
Nonetheless, an approach based on the mixture remains frequentist if non-optimal. (The chapter later attacks
the derivation of the likelihood principle by \citealp{birnbaum:1962} in Part III by Deborah Mayo that will not
be discussed in this paper as it deserves a specific paper.)

    \beq``Many seem to regard reference Bayesian theory to be a resting point until satisfactory subjective or
informative priors are available. It is hard to see how this gives strong support to the reference prior
research program."{\em --- D. Cox and D. Mayo, p.302, \EI, 2010}\enq

A section also worth commenting is (unsurprisingly!)~the one addressing the limitations of the Bayesian
alternatives (pp.298-302). The authors however dismiss straight away the personalistic approach to prior
construction by considering it fails the objectivity canons. This seems too hasty to
me, since the choice of a prior is
\begin{enumerate}
\item the choice of a reference probability measure against which to assess the
information brought by the data, not clearly less objective than picking one frequentist estimator or another,
and 
\item a personal construction of the prior can also be defended on objective grounds, based on the past
experience of the modeller. That it varies from one modeller to the next is not an indication of subjectivity per
se, simply of different past experiences. 
\end{enumerate}
Cox and Mayo then focus on reference priors, \`a la Bernardo-Berger (\citeyear{berger:bernardo:1992}), once
again pointing out the lack of uniqueness of those priors as a major flaw. While the sub-chapter agrees on the
understanding of those priors as convention or reference priors, aiming at maximising the information input from
the data, it gets stuck on the impropriety of such priors: ``if priors are not probabilities, what then is the
interpretation of a posterior?" (p.299). This seems like a strange (and unfair) comment to me:  the
interpretation of a posterior is that it is a probability distribution and this is the only mathematical
constraint one has to impose on a prior. (Which may indeed be a problem in the derivation of reference priors.)
As detailed in {\em The Bayesian Choice} among other books \citep{hartigan:1983,berger:1985}, there are many
compelling reasons to invite improper priors into the game. (And one not to, namely the difficulty with point
null hypotheses.) While I agree that the fact that some reference priors (like matching priors, whose
discussion p.302 escapes me) have good frequentist properties is not compelling within a Bayesian framework, it
seems a good enough answer to the more general criticism about the lack of objectivity: in that sense,
frequency-validated reference priors are part of the huge package of frequentist procedures and cannot be
dismissed on the basis of being Bayesian. That reference priors are possibly at odd with the likelihood
principle does not matter so much:  the shape of the sampling distribution is part of the prior information,
not of the likelihood {\em per se}. The final argument (Section 12) that Bayesian model choice requires the
preliminary derivation of ``the possible departures that might arise" (p.302) has been made at several points
in \EI~and already discussed above. Besides being in my opinion a valid working principle, i.e. selecting the
most appropriate albeit false model, this definition of well-defined alternatives is mimicked by the assumption
of ``statistics whose distribution does not depend on the model assumption" (p.302) found in the same last
paragraph.

In conclusion this (sub-)chapter by David Cox and Deborah Mayo is (as could be expected!)~a deep and thorough
treatment of the frequentist approach to the sufficiency and (weak) conditionality principle. It however fails
to convince me that there exists a ``unique and unambiguous" frequentist approach to all but the most simple
problems. At least, from reading this chapter, I cannot find a working principle that would lead me to this
single unambiguous frequentist procedure.

\subsection{Spanos' comments (Part IV)}
    \beq``It is refreshing to see Cox and Mayo give a hard-nosed statement of what scientific objectivity demands
of an account of statistics, show how it relates to frequentist statistics, and contrast that with the notion
of ``objectivity" used by O-Bayesians."{\em ---A. Spanos, p.326, \EI, 2010}\enq

The discussion by Aris Spanos is an exegesis of Part II by David Cox and Deborah Mayo: the first point in the
discussion is that the above is ``a harmonious blend of the Fisherian and N-P perspectives to weave a coherent
frequentist inductive reasoning anchored firmly on error probabilities"(p.316). The discussion by Spanos being
very much a-critical, I will rather expose here some thoughts of mine that came from reading this apology.
(Remarks about Bayesian inference are limited to some piques like the above, which only reiterates those found
earlier [and later: "the various examples Bayesians employ to make their case involve some kind of "rigging" of
the statistical model", Aris Spanos, p.325; "The Bayesian epistemology literature is filled with shadows and
illusions", Clark Glymour, p.  335] in \EI.)

The ``general frequentist principle for inductive reasoning" (p.319) at the core of Cox and Mayo's paper is
obviously the central role of the $p$-value in ``providing (strong) evidence against the null $H_0$ (for a
discrepancy from $H_0$)". Once again, I fail to see it as the epitome of a working principle in that
\begin{enumerate}
   \item it depends on the choice of a divergence measure $d(z)$, which reduces the information brought by the
data $z$;
   \item it does not articulate the level for labelling nor the consequences of finding a low $p$-value;
   \item it ignores the role of the alternative hypothesis.
\end{enumerate}
Furthermore, Spanos' discussion deals with ``the fallacy of rejection" (pp.319-320) in a rather artificial (if
common throughout \EI) way, namely by setting a buffer of discrepancy γ around the null hypothesis. While the
choice of a maximal degree of precision sounds natural to me (in the sense that a given sample size should not
allow for the discrimination between two arbitrary close values of the parameter), the fact that $\gamma$ is
{\em in fine} set by the data (so that the $p$-value is high) is fairly puzzling. If I do understand correctly,
the change from a $p$-value to a discrepancy $\gamma$ is a fine device to make the ``distance" from the null better
understood, but it has an extremely limited range of application. If I {\em do not} understand correctly, the
discrepancy $\gamma$ is fixed by the statistician and then this sounds like an extreme form of prior selection.

There is another issue I do not understand in this part, namely the meaning of the severity evaluation
probability
$$
P(d(Z) > d(z_0);\,\mu> \mu_1)
$$
as the conditioning on the range of parameter values seems impossible in a frequentist setting. This leads me to an idle and
unrelated questioning as to whether there is a solution to
$$
\sup_d \mathbb{P}_{H_0}(d(Z) \ge d(z_0))
$$
as this would be the ultimate discrepancy $\gamma$. Or whether this does not make any sense, because of the ambiguous
role of $z_0$, which somehow needs to be integrated out. (Otherwise, $d$ can be chosen so that the probability is
$1$.)

    \beq``If one renounces the likelihood, the stopping rule, and the coherence principles, marginalizes the use of
prior information as largely untrustworthy, and seek procedures with `good' error probabilistic properties
(whatever that means), what is left to render the inference Bayesian, apart from a belief (misguided in my
view) that the only way to provide an evidential account of inference is to attach probabilities to
hypotheses?"{\em ---A. Spanos, p.326, \EI, 2010}\enq

The role of conditioning on ancillary statistics is emphasized both in the paper and in the discussion. This
conditioning clearly reduces variability, however there is no reservation about the arbitrariness of such
ancillary statistics. And the fact that conditioning any further would lead to conditioning upon the whole
data, i.e.~to a Bayesian solution. I also noted a curious lack of proper logical reasoning in the argument
that, when
$$
f(z|\theta) \propto f(z|s) f(s|\theta),
$$
using the conditional ancillary distribution is enough, since ``any departure from $f(z|s)$ implies that the
overall model is false" (p.322), but not the reverse. Hence, a poor choice of $s$ may fail to detect a
departure. (Besides the fact that  fixed-dimension sufficient statistics $s$ do not exist outside exponential
families.) Similarly, Spanos expands about the case of a minimal sufficient statistic that is independent from
a maximal ancillary statistic, but such cases are quite rare and limited to exponential families [in the iid
case]. Plus, the notion of conditional ancilarity is fraught with dangers, as exposed in \cite{basu:1988}.
Still in the conditioning category, he also supports Mayo's argument against the likelihood principle being a
consequence of the sufficiency and weak conditionality principles.  However, he does not provide further
evidence against Birnbaum's result, arguing rather in favour of a conditional frequentist inference I have
nothing to complain about. (I fail to perceive the appeal of the Welch uniform example in terms of the
likelihood principle.)

\section{Conclusion}

In a overall recap, let me restate that this note about \EI~is an extended reading note and is thus far from
pretending at bringing a global and definitive Bayesian reply to the philosophical arguments raised in the
volume. Once again, this broader perspective is partly provided by \cite{gelman:shalizi:2012}. While the core
goal is of ``taking some crucial steps towards legitimating the philosophy of frequentist statistics" (p.328), 
the debate cannot escape veering at times towards a comparison with the Bayesian approach, hence generating the
above comments of mine's. From a more global perspective, reading \EI~is a worthy and fruitful exercise for 
Bayesians and frequentists alike as most chapters at least should bring new food for their
thoughts on hypothesis testing and model choice, since those inferential goals are still wide open to
improvements and rebuttals within both approaches.

\section*{ACKNOWLEDGEMENTS}
This research is supported partly by the Agence Nationale de la Recherche through the 2009-2012 grants {\sf Big
MC} and {\sf EMILE} and partly by the Institut Universitaire de France (IUF). Comments from Andrew Gelman and
discussions with Deborah Mayo are gratefully acknowledged.

\end{document}